\documentclass[twoside,slac_one]{revtex4}
\usepackage{graphicx}
\usepackage{fancyhdr}
\usepackage{amsmath} % American Mathematics Society standards
\usepackage{bm}% bold math
\usepackage{amsxtra}
\usepackage{amssymb}
\usepackage{amsthm}
\usepackage{latexsym}
\usepackage{lscape}

\def\pt{\ensuremath{p_{\mathrm{T}}}}

\def\MET{\ensuremath{E_{\mathrm{T}}^{\mathrm{miss}}}} 
\def\met{\ensuremath{E_{\mathrm{T}}^{\mathrm{miss}}}}
\newcommand{\Ztt}{\ensuremath{Z \to \tau\tau}}
\newcommand{\Mll}{\ensuremath{m_{\ell\ell}}}
\newcommand{\mz}{\ensuremath{m_{Z}}}
\def\antibar#1{\ensuremath{#1\bar{#1}}}
\def\ttbar{\antibar{t}}

\begin{document}

\title{Measurement of Single-top Quark Production with the ATLAS Detector}

\author{J. L. Holzbauer on behalf of ATLAS}
\affiliation{Department of Physics and Astronomy, Michigan State University, East Lansing, MI, USA}

\begin{abstract}
Single-top production has been studied with the ATLAS detector using 0.7$~\mathrm{fb}^{-1}$ of 2011 data recorded at 7 TeV center-of-mass energy.  The measurement of electroweak production of top-quarks allows probes of the Wtb vertex and a direct measurement of the CKM matrix element $|V_{tb}|$.  It is also expected to be sensitive to new physics such as flavor changing neutral currents or heavy W production.  The t-channel cross-section measurements are performed using both a cut-based and neural network approach, while a cut-based selection in the dilepton channel is used to derive a limit on the associated (Wt) production.  An observed cross-section of $90^{+32}_{-22}$ pb ($65^{+28}_{-19}$ pb expected) is obtained for the $t$-channel, which is consistent with the Standard Model expectation.  For the Wt production, an observed limit of $\sigma(pp\rightarrow Wt+X)~<~39.1$ pb ($40.6$ pb expected) is derived, which corresponds to about 2.5 times the Standard Model expectation.
 
\end{abstract}

\maketitle

\thispagestyle{fancy}

%%%%%%%%%%%%%%%%%%%%%%%%%%%%%%%%%%

\section{Introduction}
\begin{figure}[h]
\centering
\includegraphics[width=60mm]{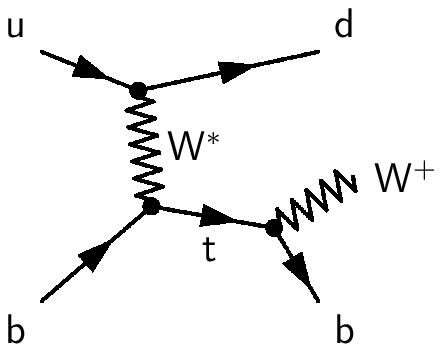}
\includegraphics[width=45mm]{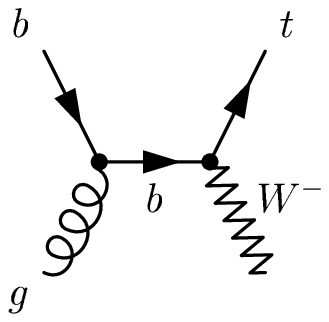}

\caption{Feynman diagrams for the $t$-channel single-top production (left) and associated single-top production, Wt (right).}
\label{fig:feynman}
\end{figure}

Single-top is electroweak top quark production and consists of three different production modes: $t$-channel, $s$-channel and Wt (associated production).  Of these channels, $t$-channel (Figure~\ref{fig:feynman}) has the largest predicted cross-section at the LHC, $64.6^{+3.3}_{-2.6}$ pb~\cite{Kidonakis:2011wy}, followed by Wt (Figure~\ref{fig:feynman}) with $15.7\pm1.4$ pb~\cite{Kidonakis:2010ux}, and s-channel with $4.6\pm0.3$ pb~\cite{Kidonakis:2010tc}.  This is in contrast to the Tevatron, where the Wt production is the smallest mode~\cite{Kidonakis:2006tev}.  The $t$-channel cross-section is about 30 times larger at the LHC than at the Tevatron, which will give the LHC experiments the opportunity to measure its cross-section with better precision.  Single-top production allows tests of the $W-t-b$ coupling via direct measurements of the CKM matrix element $|V_{tb}|$~\cite{CKM1,CKM2} as well as searches for deviations in the value of this coupling due to flavor changing neutral currents or heavy bosons~\cite{Tait:2000sh}. 

Single-top production was first observed at the Tevatron in 2009 by the D0~\cite{Abazov:2009ii} and CDF~\cite{Aaltonen:2009jj} collaborations and a combined $s$-channel and $t$-channel result was also reported~\cite{Group:2009qk}.  Later the $t$-channel production mode of single-top alone was observed by D0~\cite{Abazov:2011rz}.  Measurements of single-top t-channel production have also been reported at the LHC by CMS~\cite{Chatrchyan:2011vp} and by ATLAS~\cite{ATLAS-CONF-2011-027, ATLAS-CONF-2011-088}, which also reported the first limits on the Wt single-top process (associated production)~\cite{ATLAS-CONF-2011-027}.  Here, we will discuss the updated ATLAS results with 0.7$~\mathrm{fb}^{-1}$ for both the $t$-channel~\cite{ATLAS-CONF-2011-101} and Wt-channel~\cite{ATLAS-CONF-2011-104} analyses.  

\section{Single-top t-channel Analysis}
The $t$-channel production cross-section is measured using both a cut-based and a neural network approach.  In both cases the analyses start by reducing the large multijet and W+jets backgrounds through a common event selection which selects for the t-channel final state.  Events with at least 2 jets with transverse momentum of $\pt > 25~\mathrm{GeV}$ and $|\eta| < 4.5$ are selected, and the analysis is performed using only the 2-jet or 3-jet multiplicity bins.  Among the selected jets, exactly 1 jet must be b-tagged using a secondary vertex b-tagging algorithm with a 50\% b-tagging efficiency and a rejection of 270 for light-quarks.  Events with exactly 1 isolated, central ($|\eta| < 2.5$) muon or electron with $\pt > 25~\mathrm{GeV}$ are selected and where each of these leptons must be associated with a corresponding lepton trigger.  The events are also required to have missing transverse energy, $\met > 25~\mathrm{GeV}$ to account for the undetected neutrino from the W decay.  Events where the W boson decays into hadrons are not considered in these analyses.  Finally, $\met + M_{\mathrm{T}}(W) > 60~\mathrm{GeV}$ is required to further reduce multijet contributions, where $M_{\mathrm{T}}(W)$ is the reconstructed transverse mass of the W determined from the lepton and \met.  After all of these selections, the signal yield divided by the background yield (S/B) is approximately 0.1.

\begin{figure}[h]
\centering
\includegraphics[width=80mm]{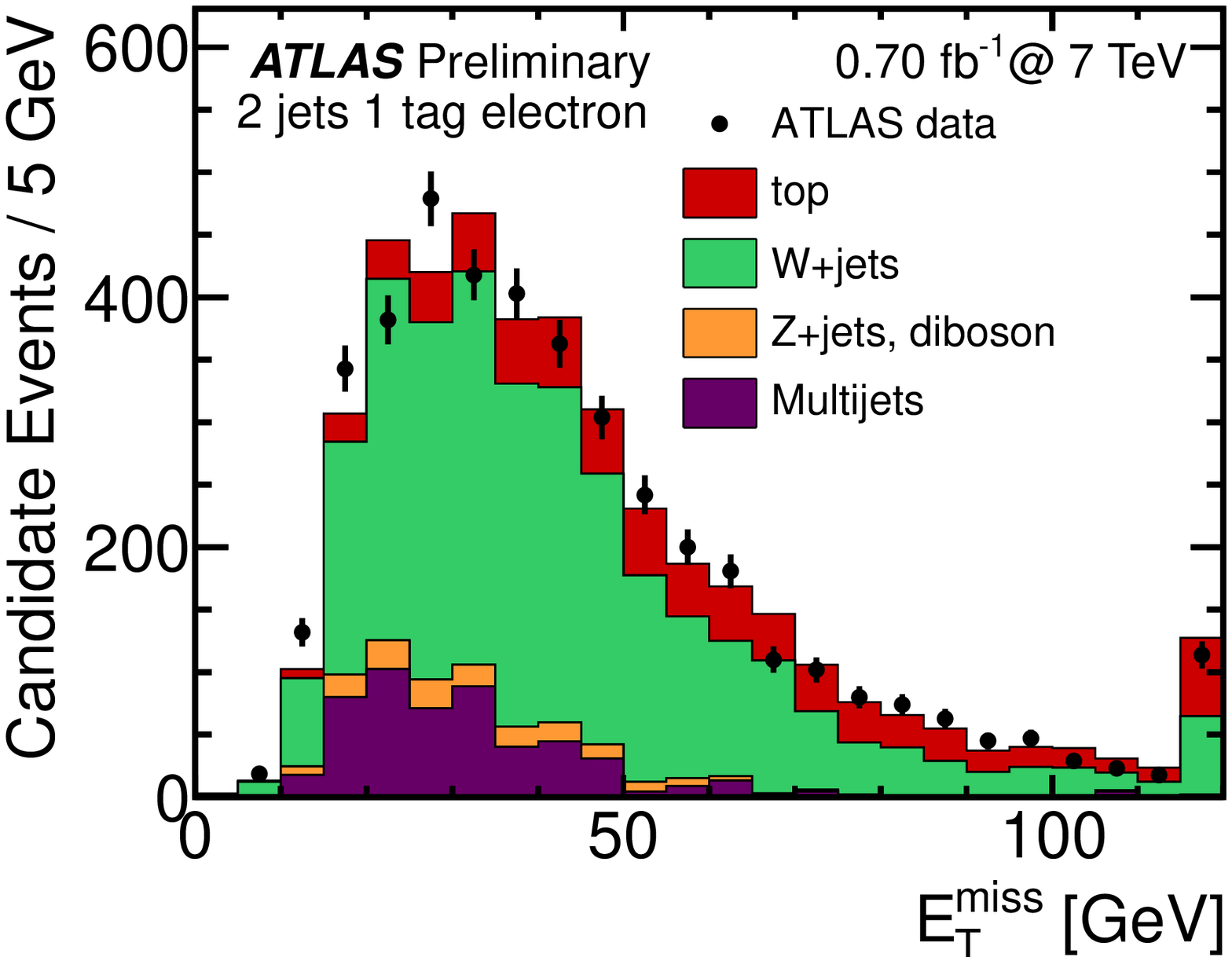}
\includegraphics[width=80mm]{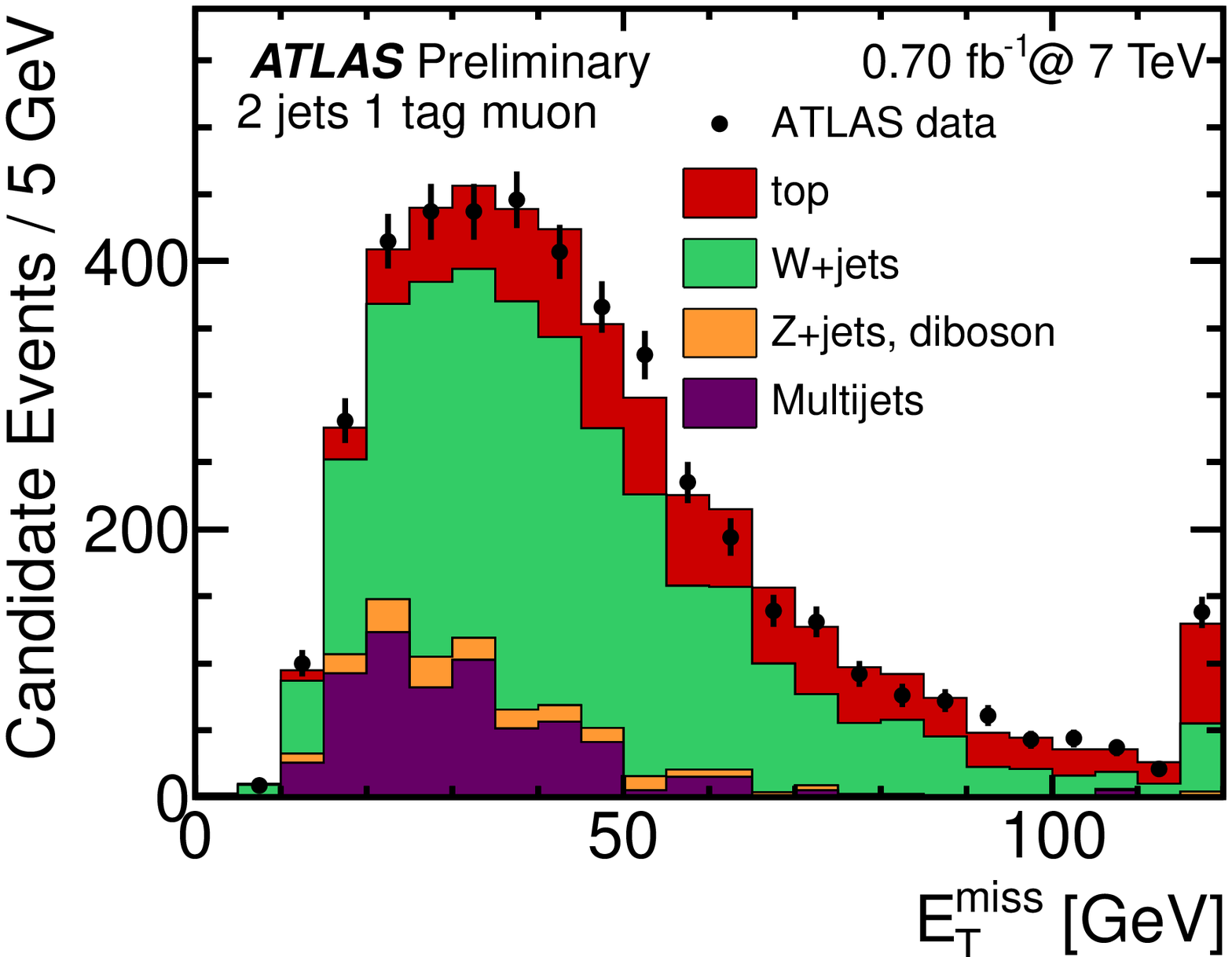}
\caption{\met~ distributions used in the multijet determination for the electron selection (left) and muon selection (right)}
\label{fig:met}
\end{figure}

A few different approaches are used to determine the background normalizations and shapes.  For the \ttbar, Wt, $s$-channel, diboson, and Z+jets backgrounds the theoretical cross-sections are used.  For the multijet estimate a ``jet-electron'' method is used, where the normal electron trigger is replaced by a jet trigger.  The normalization is determined by a binned maximum likelihood fit to the \met~ distribution in data~\cite{Aad:2010ey_sgtop}, as shown in Figure~\ref{fig:met}.  This method is used in both the electron and muon channels.  

\begin{figure}[h]
\centering
\includegraphics[width=80mm]{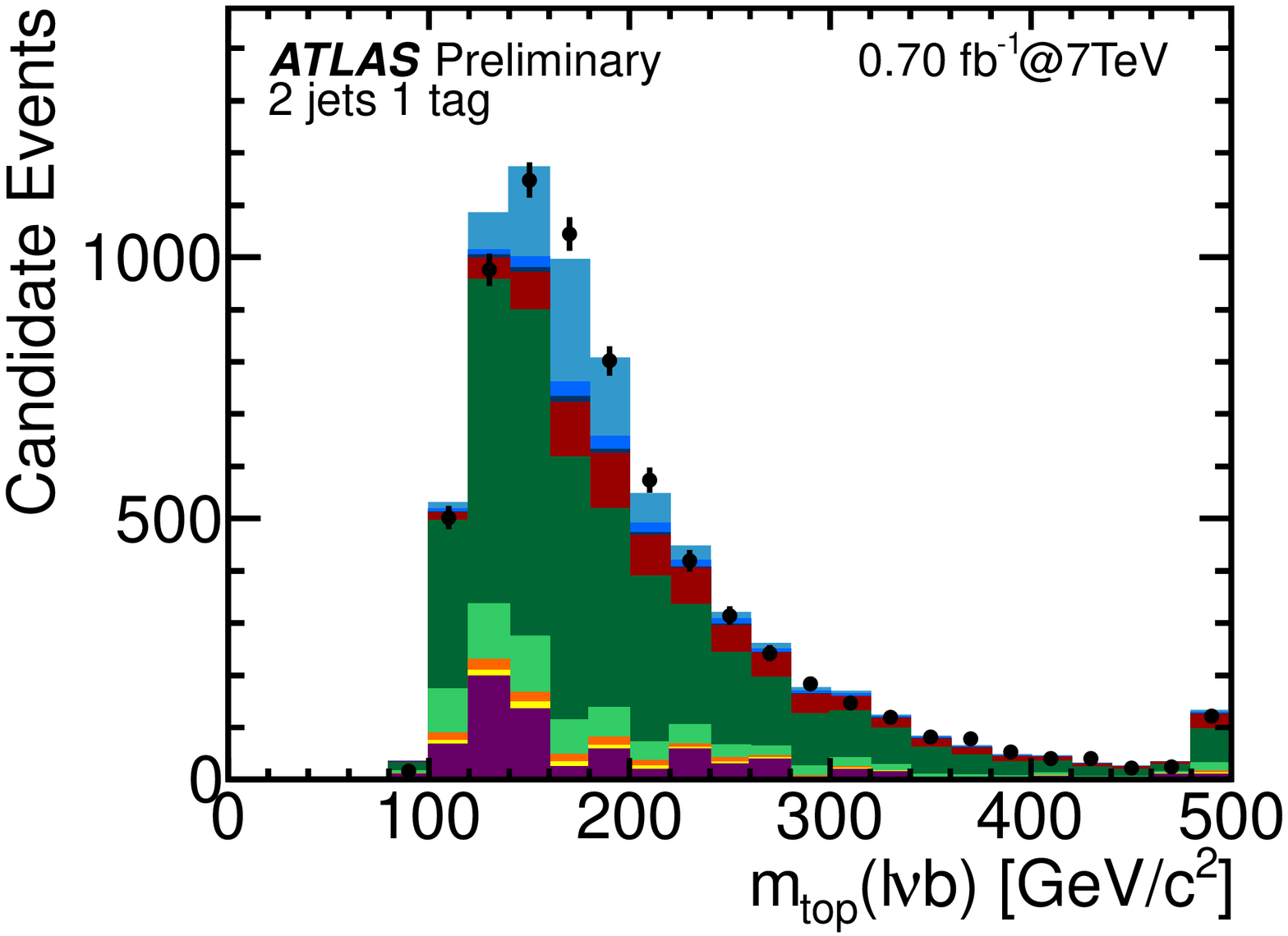}
\includegraphics[width=80mm]{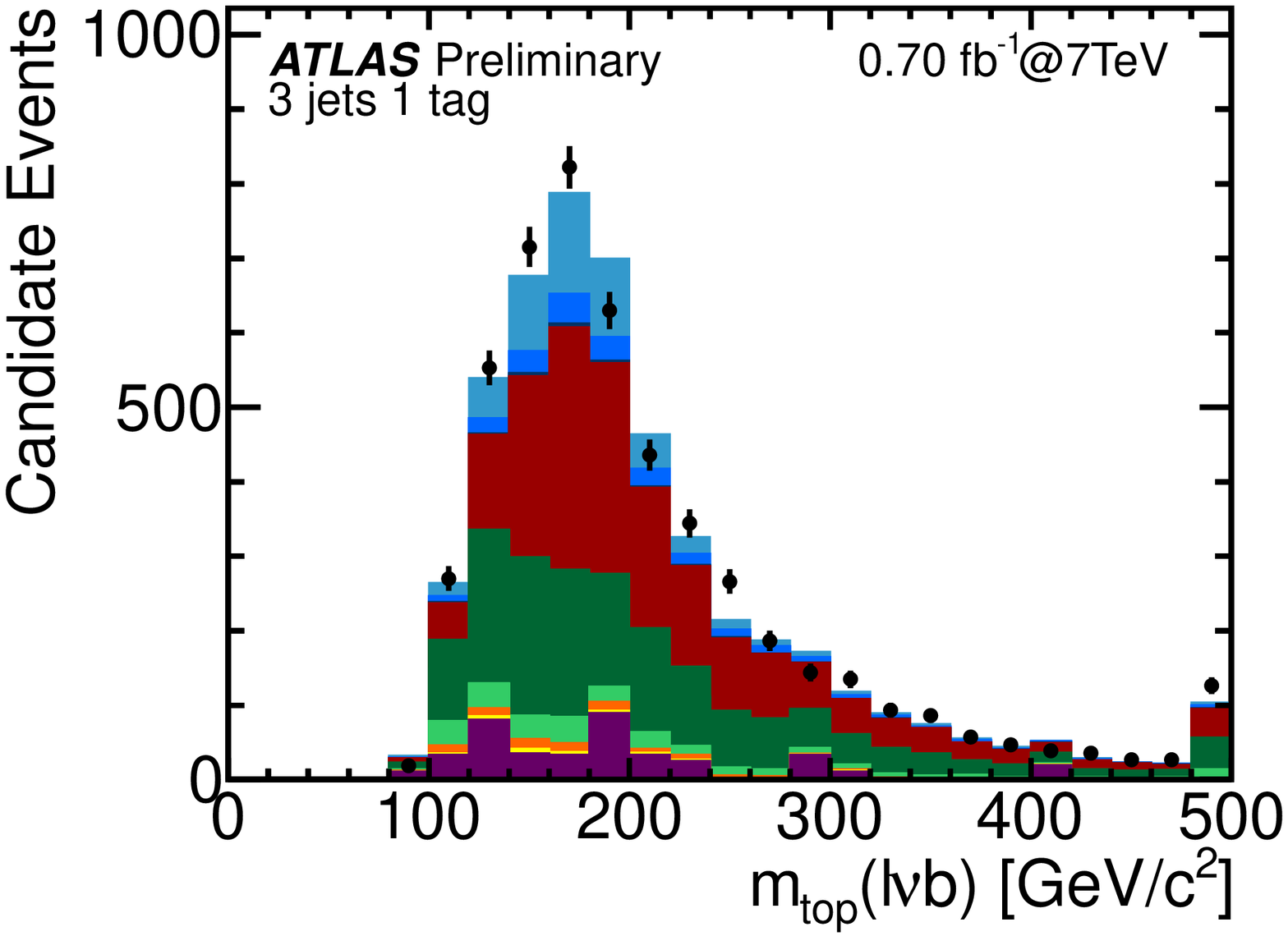}
\includegraphics[width=80mm]{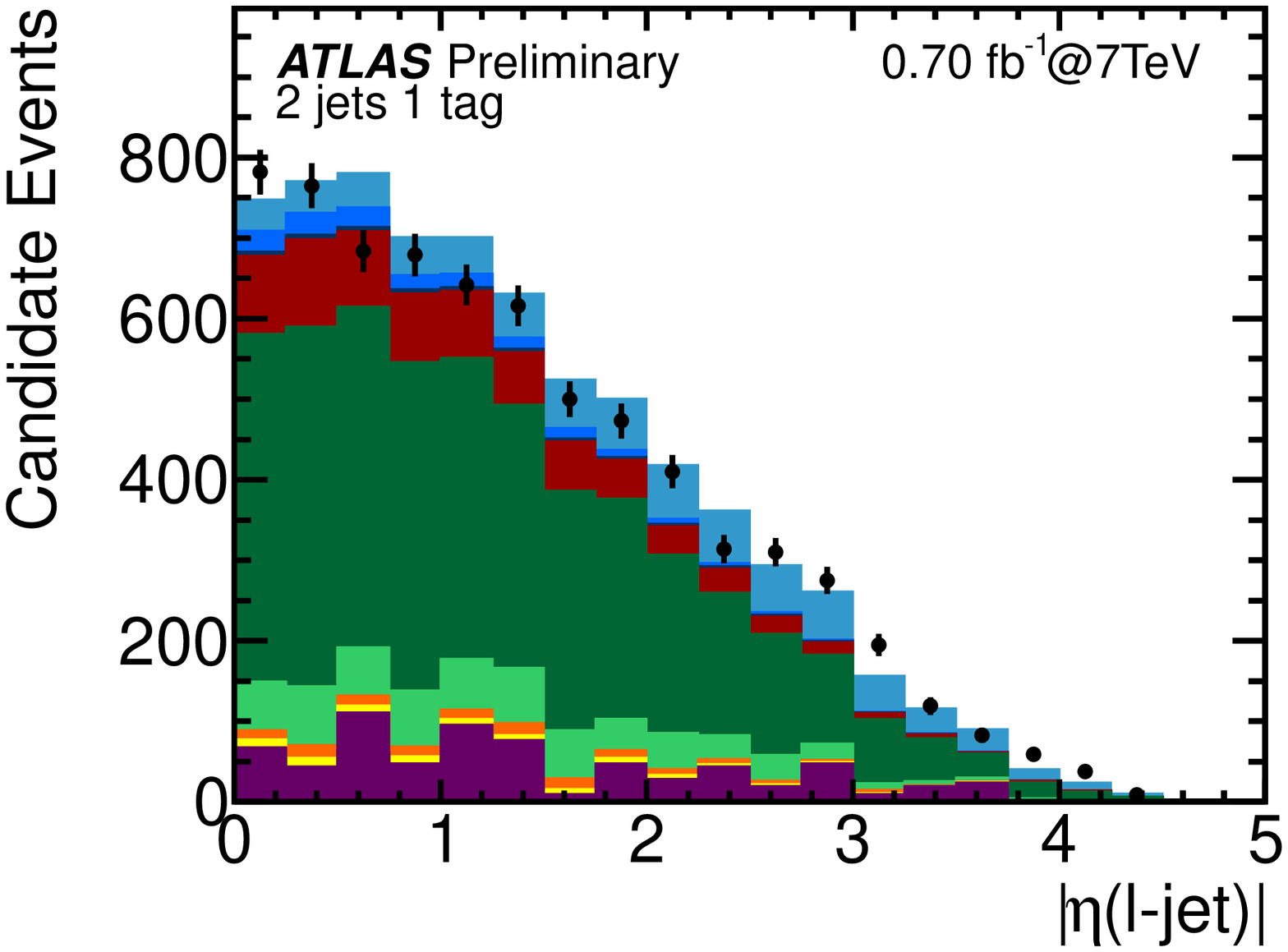}
\includegraphics[width=60mm]{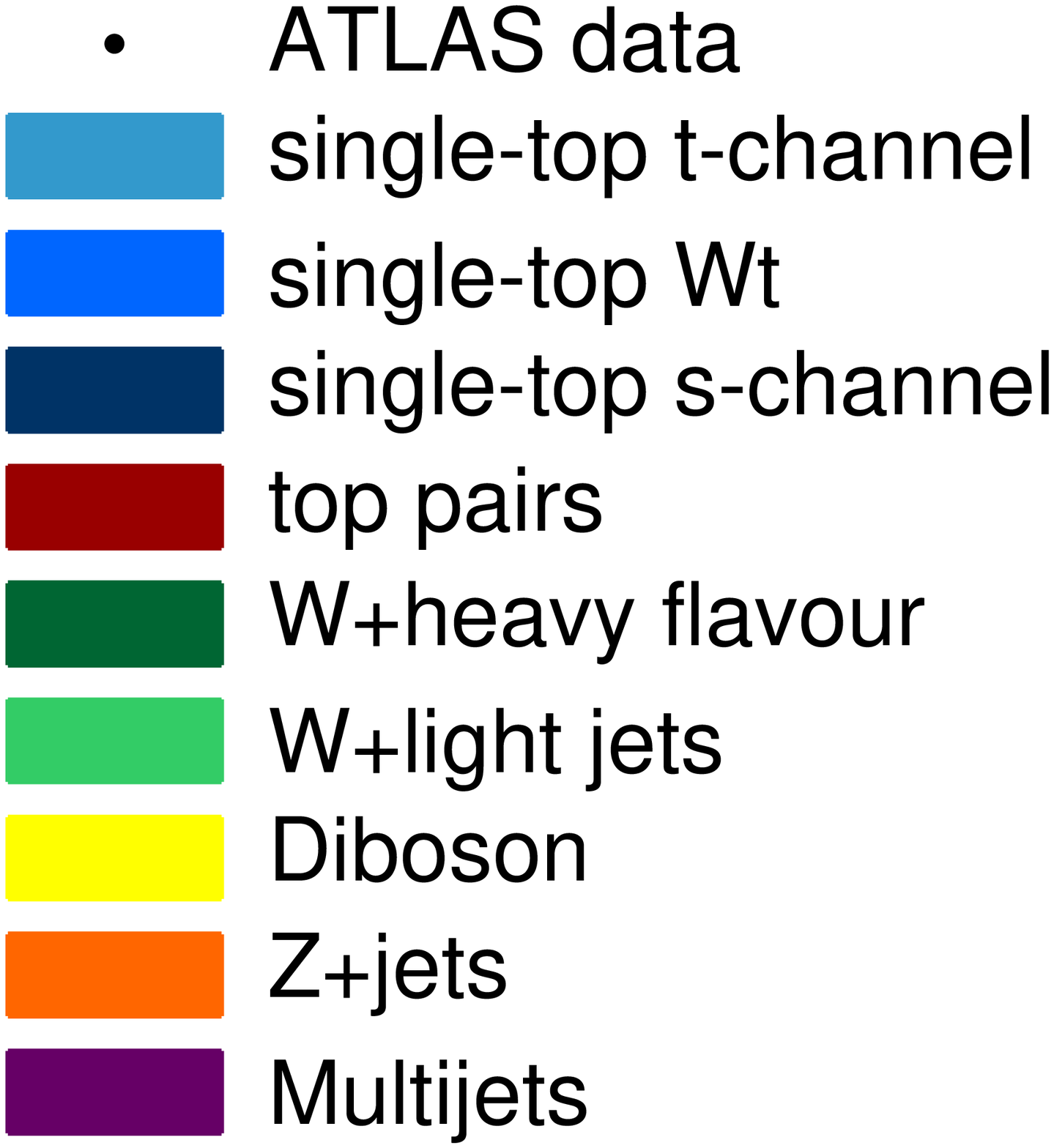}
\caption{Two important variables for the $t$-channel analyses are shown after the initial selection, including reconstructed top mass distribution after initial event selection for the two jet selection (top left), three jet selection (top right), untagged jet $\eta$ for the two jet selection (bottom left), and legend (bottom right).  The $t$-channel signal contribution is normalized to the measured combined cut-based analysis $t$-channel cross section.  Positive and negative lepton channels are combined in these distributions.}
\label{fig:topmasspresel}
\end{figure}

For the dominant W+jets backround there are two different methods employed to determine the normalization and heavy flavor fractions from data.  In the cut-based analysis, the overall normalization is determined from the 2 jet sample after the intial selection (tag) but without the b-tagged jet selection (pretag).  The heavy flavor composition of the W+light jets, Wc+jets, and the combined Wc$\bar{c}$+jets and Wb$\bar{b}$+jets proportions are determined using three orthogonal off-signal regions: 1 jet tag, 2 jet pretag, and 2 jet tag (excluding events selected by the cut-based analysis).  The neural network analysis uses a fit of the neural network output distribution to determine the W+jets normalization and heavy flavor composition.  After the background estimates and initial event selection there is good agreement between the data and signal plus background model, as seen in the variable distributions in Figure~\ref{fig:topmasspresel}.

The cut-based analysis requires four selections in addition to the standard event selection.  These are $|\eta({\rm l-jet})|>2.0$, $H_{\mathrm{T}}>210~\mathrm{GeV}$, $150~\mathrm{GeV}<M_{l\nu b}<190~\mathrm{GeV}$ and $\Delta\eta({\rm b-jet}, {\rm l-jet}) > 1.0$.  Here, l-jet refers to the untagged jet with the largest \pt~ value and b-jet refers to the b-tagged jet.  Additionally, $H_{\mathrm{T}}$ is sum of the transverse momentum of the leading two jets in \pt, the lepton \pt, and \met~ while $M_{l\nu b}$ is the reconstructed top quark mass.  This analysis uses four different channels (where each channel can contain both electron and muon channel events): 2 and 3 jet events with positively or negatively charged leptons.  This channel division makes use of the expected lepton charge asymmetry from the proton-proton collisions due to the excess of valence u-quarks compared to valence d-quarks.  This is different than a proton-antiproton collider, where no asymmetry would be expected.  

\begin{figure}[h]
\centering
\includegraphics[width=80mm]{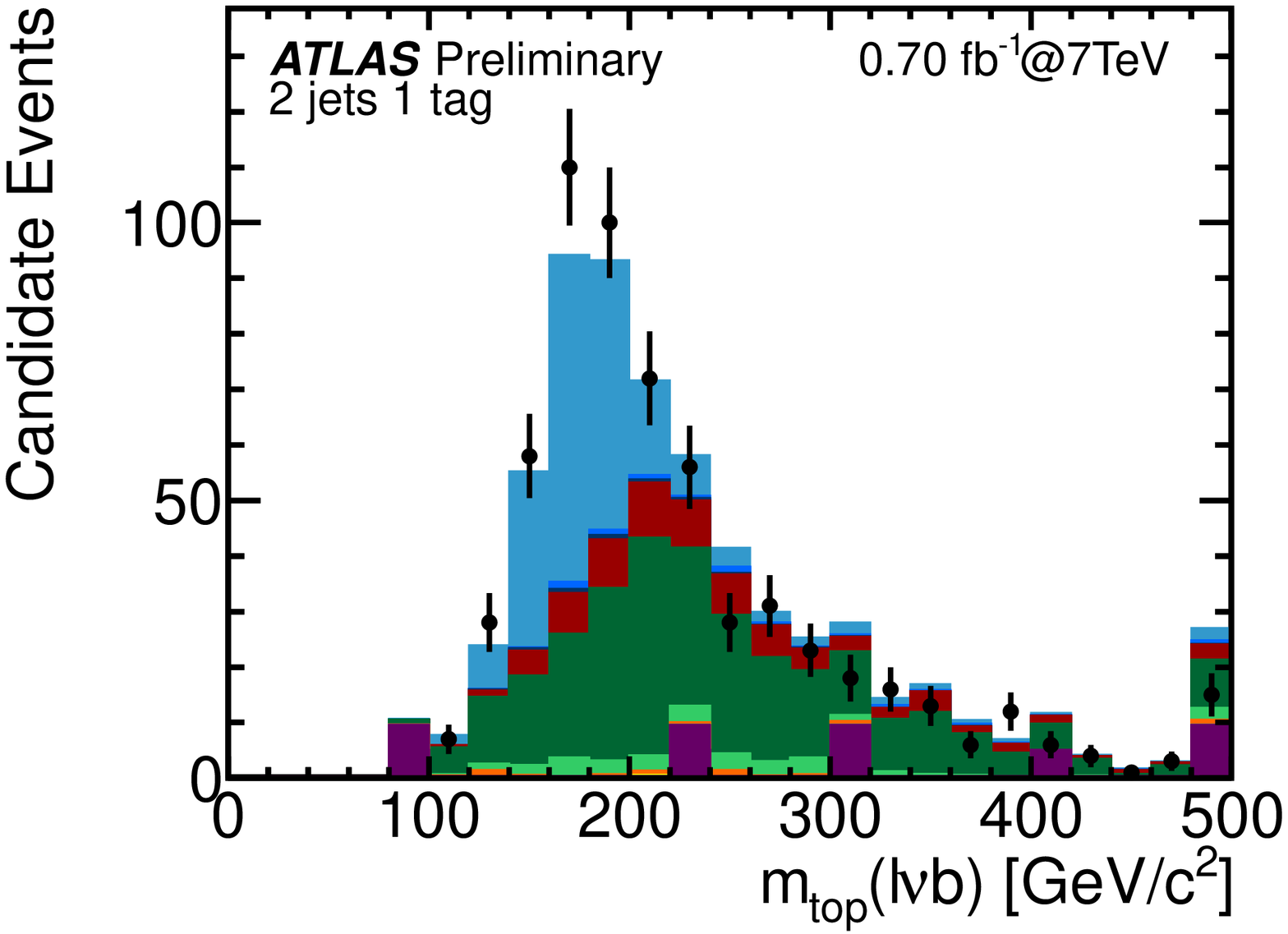}
\includegraphics[width=80mm]{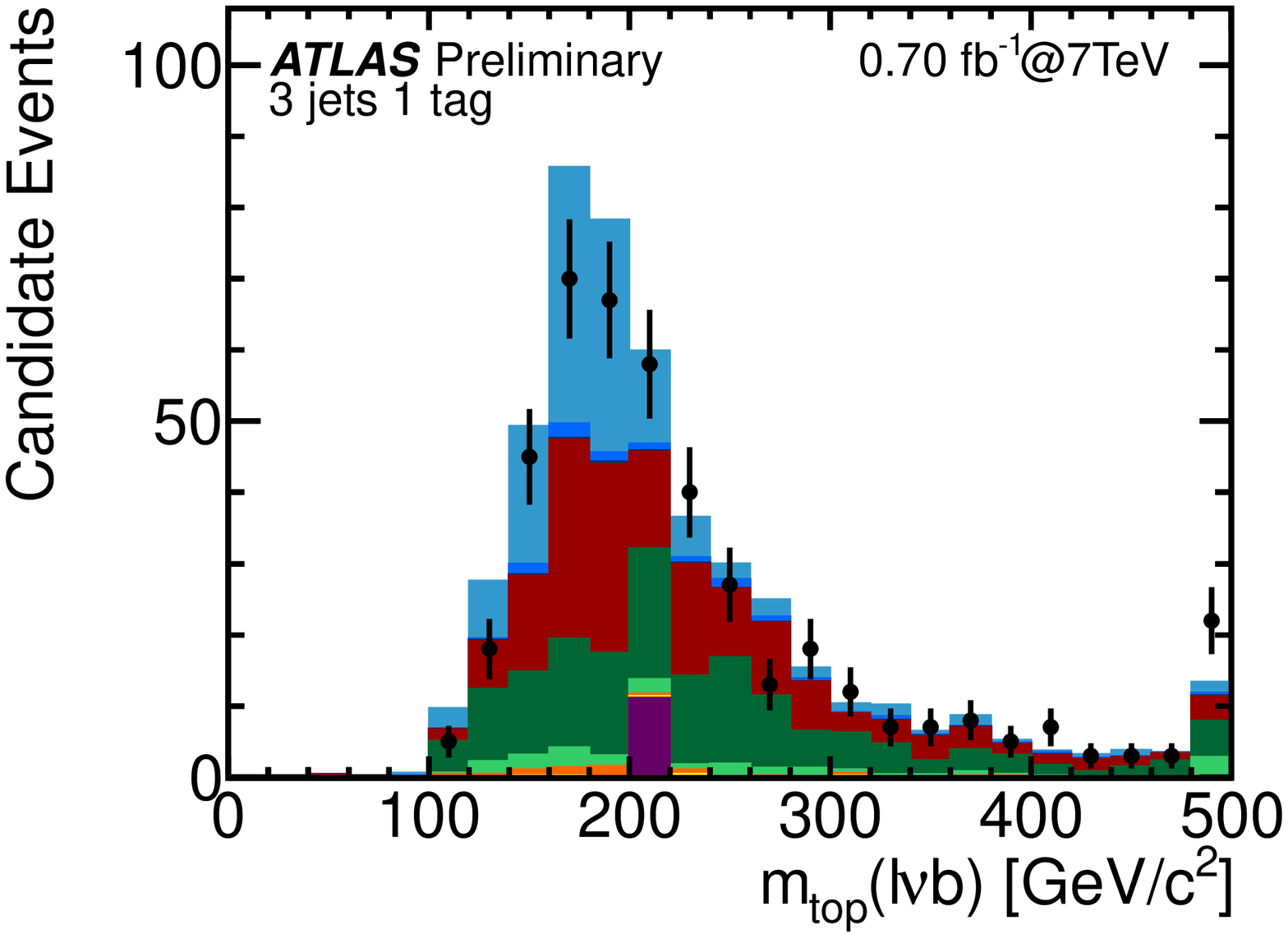}
\includegraphics[width=60mm]{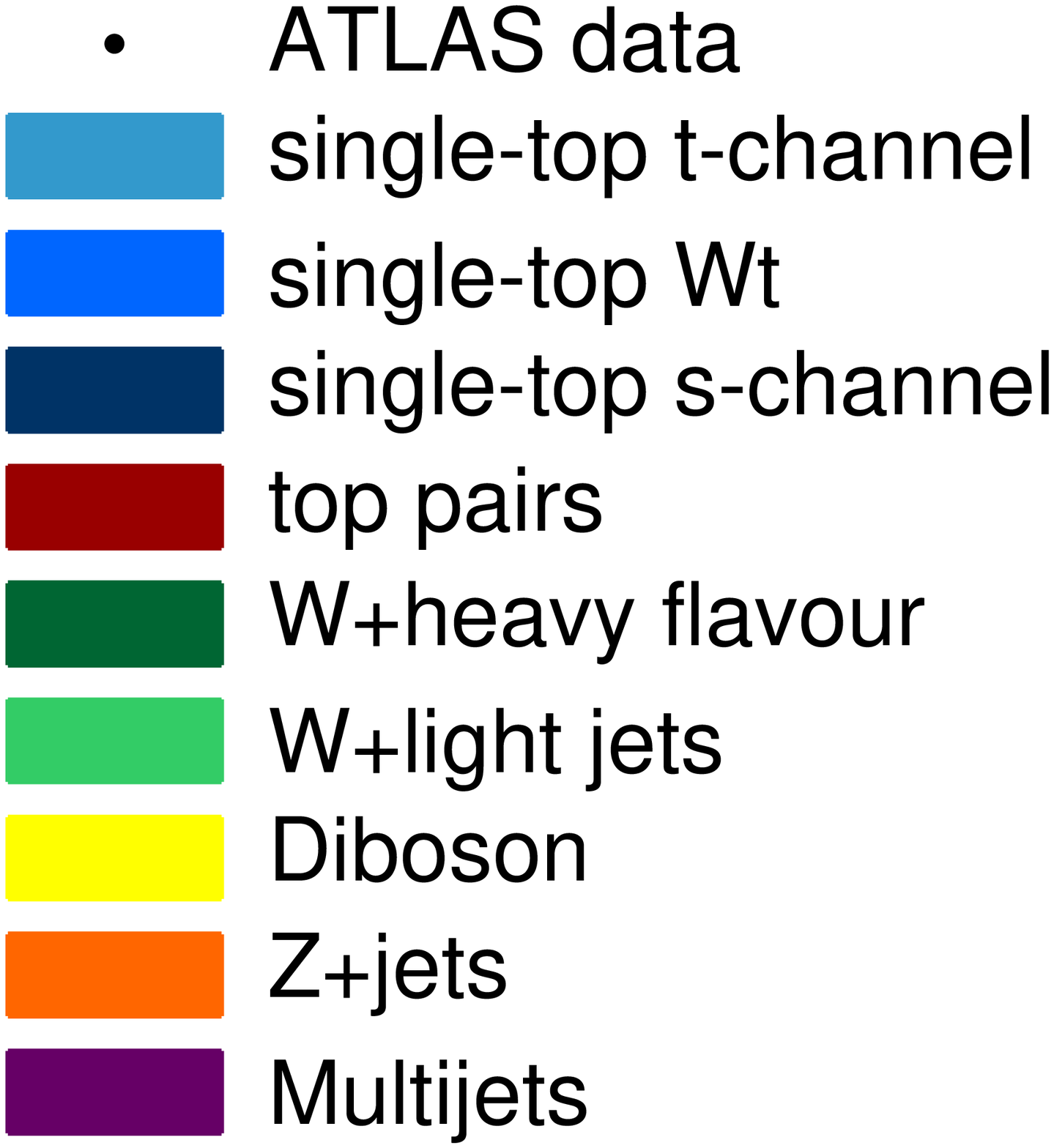}

\caption{Reconstructed top mass distribution after all additional cut-based analysis selections except the reconstructed top mass selection for the two jet selection (top left) and three jet selection (top right) with the legend (bottom). The $t$-channel signal contribution is normalized to the measured combined cut-based analysis $t$-channel cross section.}
\label{fig:topmassafterpresel}
\end{figure}

After the event selection the best channel is the 2 jet events with positively charged leptons channel followed by the 3 jet events with positively charged leptons channel.  These are the best channels particularly due to their larger proportion of signal versus \ttbar~ in the positively charged lepton channel, due to \ttbar~ production being symmetric in the lepton charge channels.  Also, there is a lower proportion of 2 jet \ttbar~ events than 3 jet \ttbar~ events, as the typical \ttbar~ event has 4 jets.  The cut-based selections are particularly helpful in reducing the other large background, W+jets, which is asymmetric in lepton charge, like the signal, and appears more often in the 2 jet bin than the 3 jet bin.  Specifically, after all cut-based analysis selections in the 2 jet (3 jet) channel with a postively charged lepton about 94 events (83 events) are expected overall with an expected signal yield of about 52 events (33 events) giving an expected S/B ratio of 1.2 (0.7), as shown in Table~\ref{tab:SB}.  This improvement can also be seen for 2 and 3 jet events with the positive and negative lepton channels combined in Figure~\ref{fig:topmassafterpresel}, which shows the distributions in Figure~\ref{fig:topmasspresel} but with the additional cut-based analysis selections (except the reconstructed top mass selection) applied.
\begin{table*}[t]
\begin{center}
\caption{Yields for signal and total signal plus background model, as well as the S/B ratio and data totals.  The cut-based analysis yields are after all of the cut-based selections.  The neural network analysis yields are the initial selection yields for events in the distribution the analysis fits.}
\begin{tabular}{|l|c|c|c|c|c|c|}
\hline 
        &\multicolumn{2}{c|}{\textbf{Cut-based 2 Jets}} &\multicolumn{2}{c|} {\textbf{Cut-based 3 Jets}}  &\multicolumn{1}{c|} {\textbf{Neural Network 2 Jets}}  \\
\hline 
        \textbf{Process} & \textbf{Lepton} + & \textbf{Lepton -}  & \textbf{Lepton +} & \textbf{Lepton -}  & \textbf{All}\\
\hline  single-top $t$-channel&51.8 $\pm$ 16.4 &   23.7  $\pm$ 6.5  &   33.0  $\pm$ 7.0 &   16.3  $\pm$ 4.8 & 900 $\pm$ 60 \\
\hline  TOTAL Expected &  94.1  $\pm$ 18.4 &   50.2  $\pm$ 8.5 &   82.6  $\pm$ 12.7 &    57.9  $\pm$ 10.1  & 6950  $\pm$ 880 \\
\hline  S/B            & 1.23 & 0.89 & 0.67 & 0.39  & 0.13 \\
 \hline DATA           & 118 & 68 & 74 & 60 & 6953 \\
\hline 
 \end{tabular}
\label{tab:SB}
\end{center}
\end{table*}

\begin{figure}[h]
\centering
\includegraphics[width=80mm]{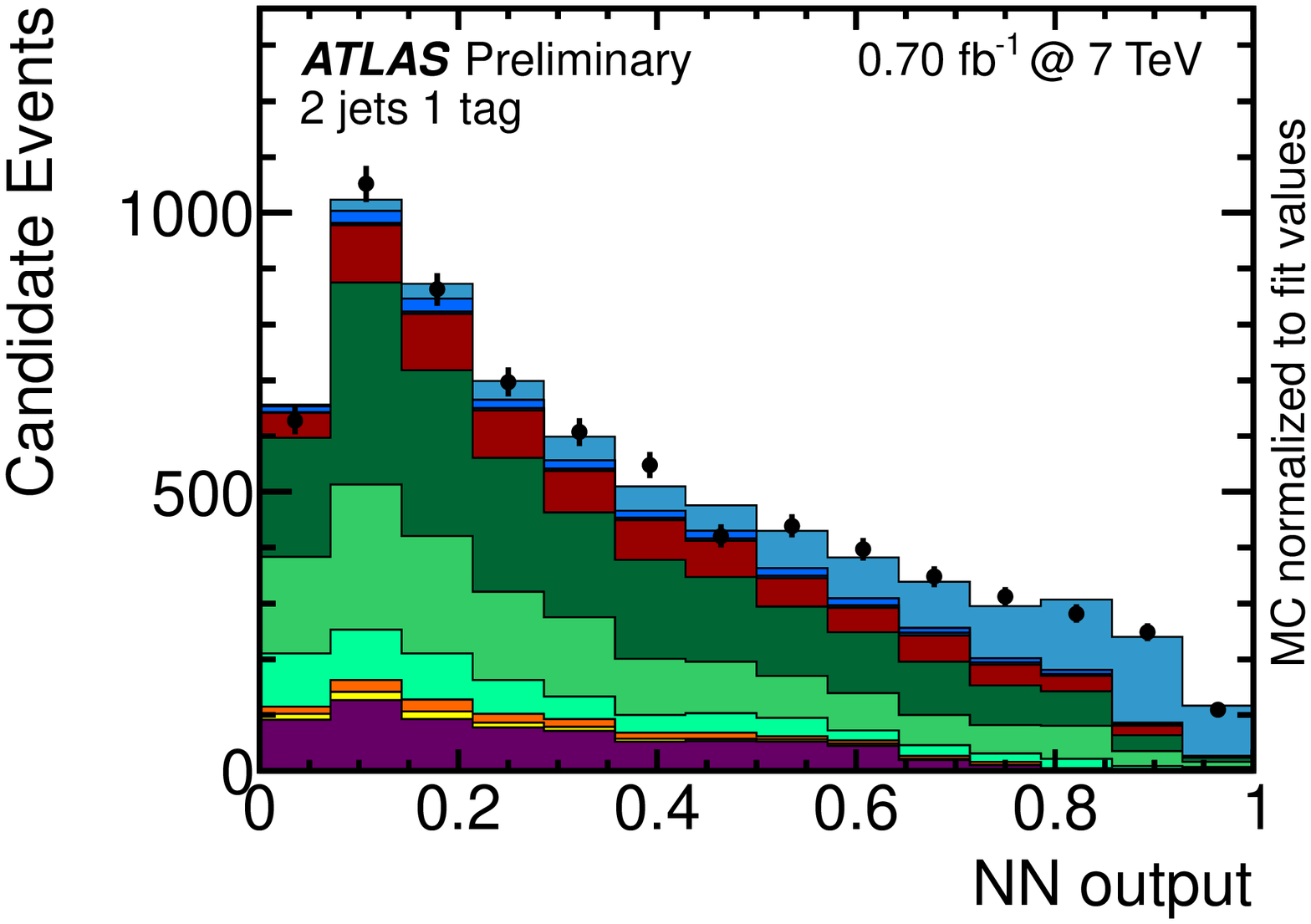}
\includegraphics[width=80mm]{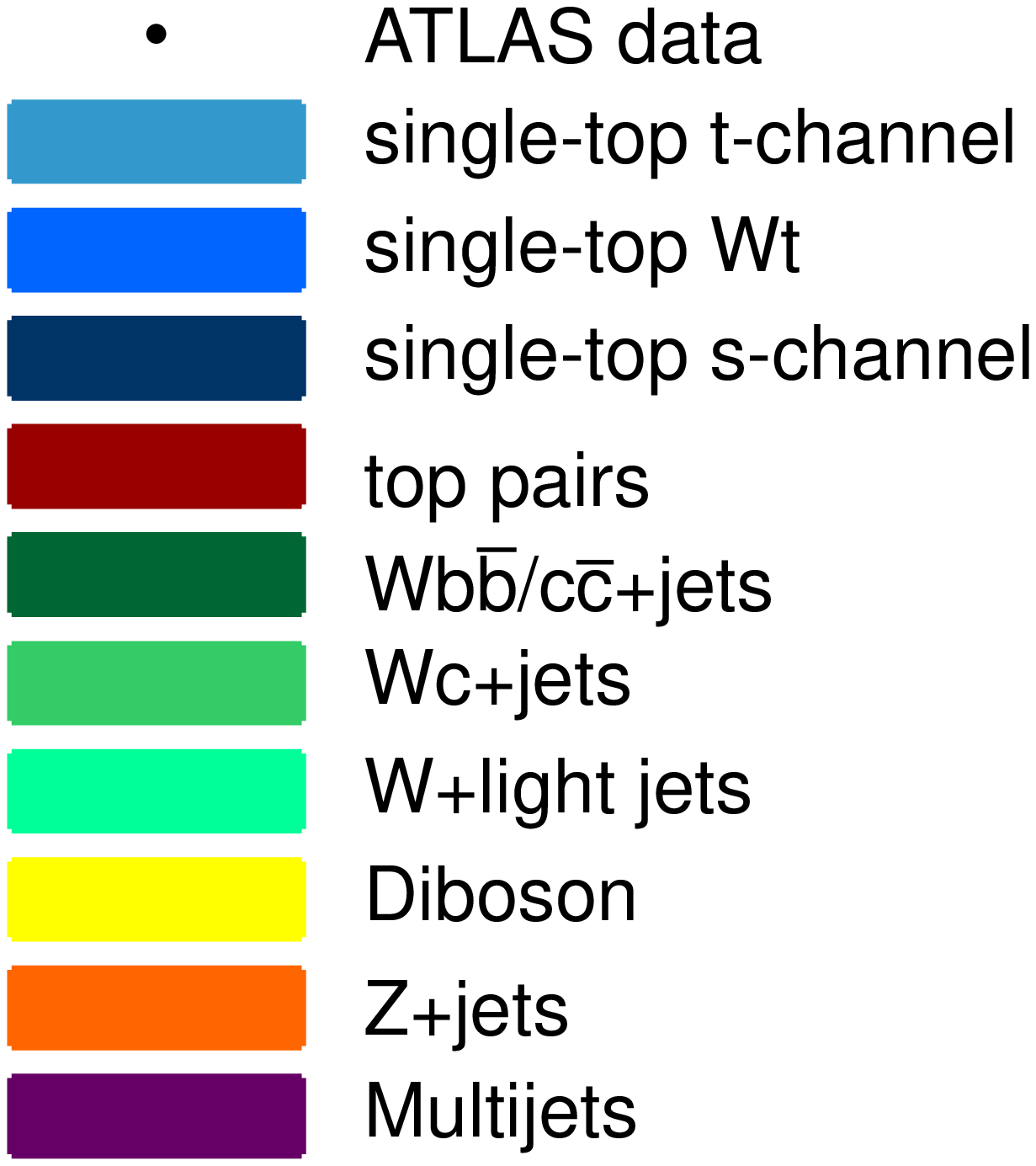}
\caption{Neural network output distribution after initial event selection.}
\label{fig:NN}
\end{figure}

The neural network analysis uses the NeuroBayes~\cite{feindt-2004,Feindt:2006pm} program with 13 variables, including the variables used in the cut-based analysis and lepton charge, used in its channel division.  The other 8 variables include the invariant mass of the two jets, $M_{\mathrm{T}}(W)$, the lepton $\eta$ and \pt, \met, the \pt~ of the untagged jet, the mass of the b-tagged jet, and $\delta\eta$ between the b-tagged jet and reconstructed W boson.  This analysis combines these 13 variables including their correlations into one discriminant.  The output distribution is then fit with a binned likelihood fit to extract the cross-section.  Unlike the cut-based analysis, no additional cuts are applied after the initial event selection and this analysis only uses two jet events.  The event yields for this selection can be seen in Table~\ref{tab:SB} along with the expected S/B ratio for the intial selection.  The neural network output distribution used in the fit can be seen in Figure~\ref{fig:NN} and the agreement between data and signal plus background model in this distribution is good.

The $t$-channel cross-sections are extracted using a profile likelihood technique for the cut-based analysis (CB) and a maximum likelihood fit for the neural network analysis (NN).  The individual observed results are $\sigma_{t}= 102^{+40}_{-30}$ pb (CB 2 jets), $\sigma_{t}= 105^{+37}_{-31}$ pb (NN), and  $\sigma_{t}=50^{+34}_{-27}$ pb (CB 3 jets), all of which are consistent within at least two standard deviations.  

The baseline result is the combination of the two and three jet channels from the cut-based analysis, $\sigma_{t}= 90^{+9}_{-9}(\mathrm{stat})\,^{+31}_{-20}(\mathrm{syst})= 90^{+32}_{-22}$ pb observed,  $\sigma_{t}^\mathrm{exp}=65^{+28}_{-19}$ pb expected.  The observed cross-section is consistent with the predicted standard model $t$-channel cross-section within about 1.1 standard deviations.  The uncertainty of this measurement is dominated by systematic uncertainties, with b-tagging, jet energy scale, and initial/final state radiation being particularly large contributors to the cross-section uncertainty.

\section{Wt Single-top Analysis}
\begin{figure}[h]
\centering
\includegraphics[width=80mm]{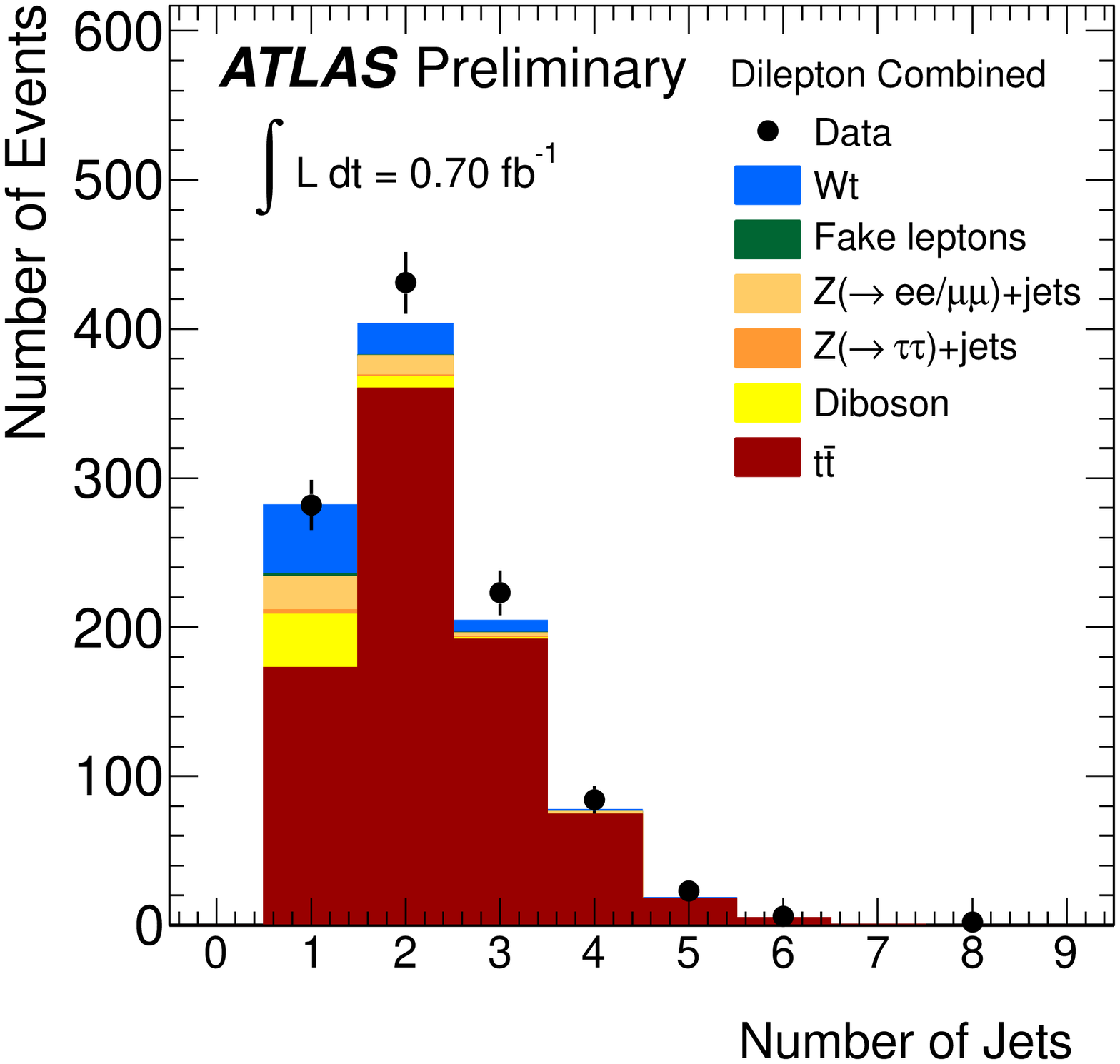}
\includegraphics[width=80mm]{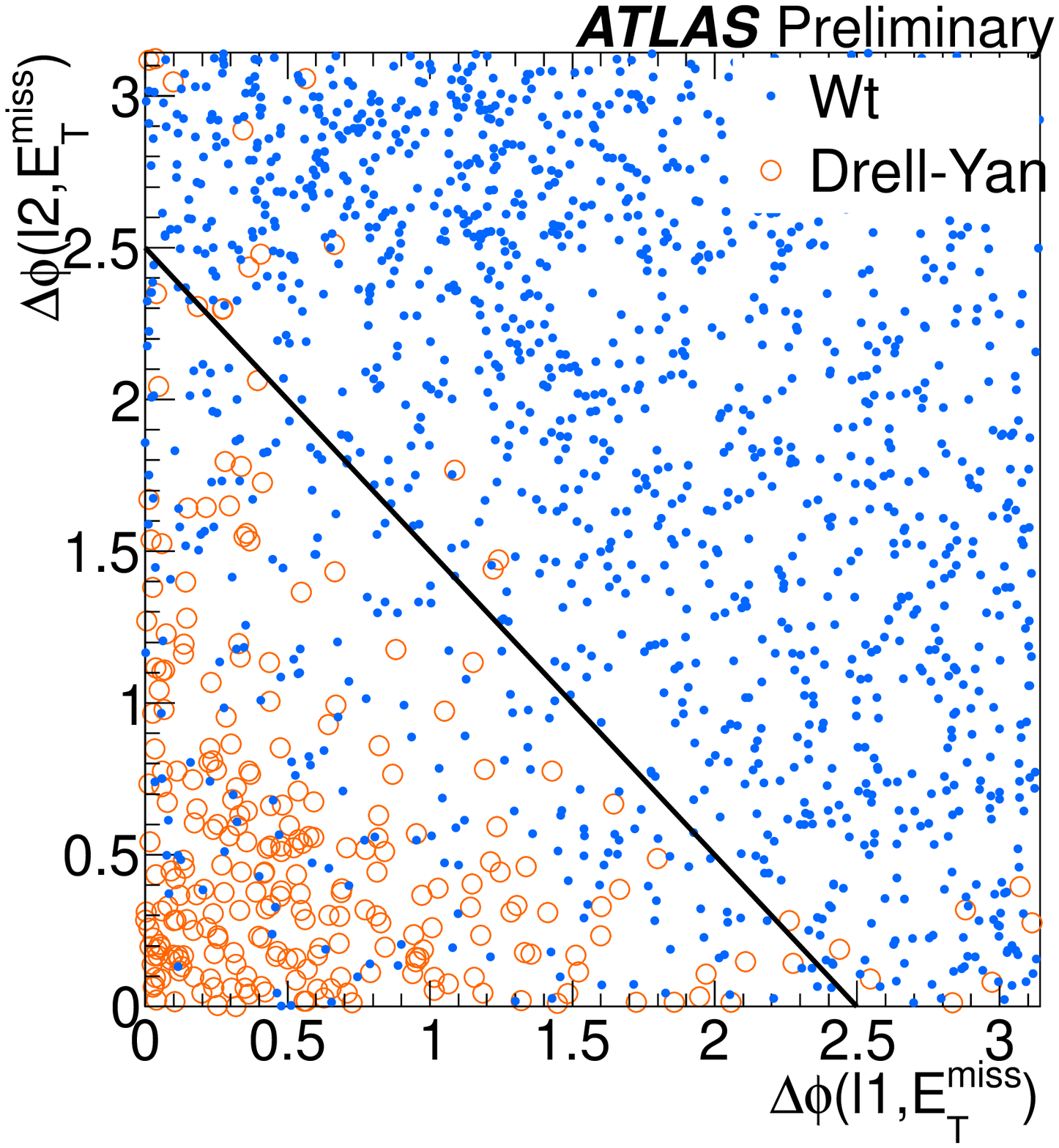}
\caption{Number of jets after intial event selection (left) and \Ztt~ selection distribution (right).}
\label{fig:WtSel}
\end{figure}

The Wt analysis uses a cut-based strategy to select the dilepton Wt single-top production signal (where both the W's decay into a lepton and neutrino).  There are three different dilepton analysis channels: electron-electron (ee), electron-muon (e$\mu$), and muon-muon ($\mu\mu$).  In all three channels, at least one jet is required with $\pt > 30~\mathrm{GeV}$, as well as exactly two isolated, triggered, central muons or electrons with $\pt > 25~\mathrm{GeV}$ and corresponding $\met > 50~\mathrm{GeV}$.  Additionally, two selection criteria are used to reduce further specific backgrounds.  The sum of the angular differences in phi between the leptons and \met~ direction is used to reduce the \Ztt~ contamination, $\Delta\phi(l_1,\MET)+\Delta\phi(l_2,\MET) > 2.5$ (see Figure~\ref{fig:WtSel}).  The other selection is a requirement on the reconstructed dilepton invariant mass which is used to remove events coming from Z boson decays, $|\Mll - \mz| > 10~\mathrm{GeV}$, and is applied to events where both leptons are electrons or both are muons.  Here, $\Mll$ refers to the reconstructed mass of the two leptons and $\mz$ refers to the Z boson mass.

The background cross-section normalizations are all estimated using data driven techniques except for the diboson background, where the theoretical cross-section is used.  The other backgrounds are fakes (backgrounds where a lepton is not a real lepton, such as multijets and W+jets processes), Drell Yan, \Ztt, and \ttbar~ (dilepton final state), which is the largest background in the final analysis due to its real leptons and top quarks.  These four background processes are estimated with different data-based techniques.  The Drell Yan cross-section normalization is estimated using a so-called ABCDEF method.  This method use two uncorrelated variables, \met~ and \Mll, and forms six different kinematic regions% as shown in Figure~\ref{fig:drellyan}
, where regions A and C are signal regions.  Non-Drell Yan events in the off-signal regions are subtracted and correlations between these two variables are corrected.  

Fakes are estimated using a matrix method~\cite{Aad:2010ey_sgtop}.  Here a four by four matrix is formed relating tight (T) and loose (L) lepton combinations (TT, TL, LT, LL) to their real (R) or fake (F) classifications (RR, RF, FR, FF).  Both leptons in the initial selection are required to be tight, so TT is the signal selection region.  Samples with Z boson decaying to two leptons are used to determine the real efficiencies, while the efficiency for a loose lepton to be identified as a tight lepton is determined with a sample containing a single loose lepton and \met~ below 10 GeV.  This matrix is then used to derive the number of fake events in the TT selected sample. 

The \ttbar~ background is estimated using a \ttbar~ dominated region orthogonal to the signal region defined as events with at least 2 jets (see Figure~\ref{fig:WtSel}).  The non-\ttbar~ events in this region are subtracted from the data and the result is compared to the expected value to determine a scale factor.  This result is then propagated into the signal region.  Systematic uncertainties for this measurement are estimated and the correlations between these uncertainties and those in the cross-section measurement are accounted for.  The \Ztt~ is estimated using a \Ztt~ dominated region orthogonal to the signal region, $\Delta\phi(l_1,\MET)+\Delta\phi(l_2,\MET) < 2.5$ (see Figure~\ref{fig:WtSel}).  Here, non-\Ztt~ events are subtracted and the result is propagated to the signal region.

\begin{figure}[h]
\centering
\includegraphics[width=80mm]{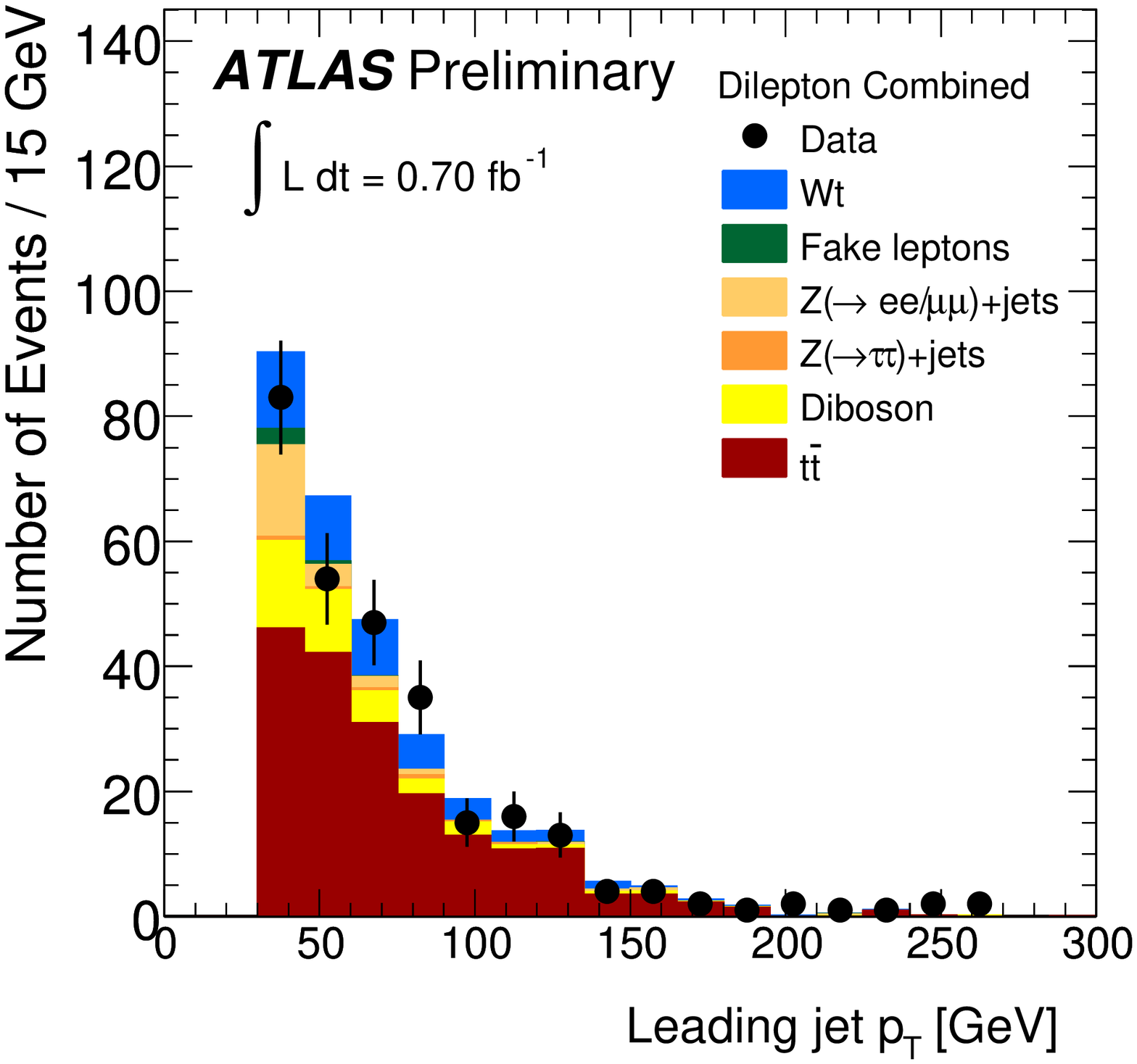}
\includegraphics[width=80mm]{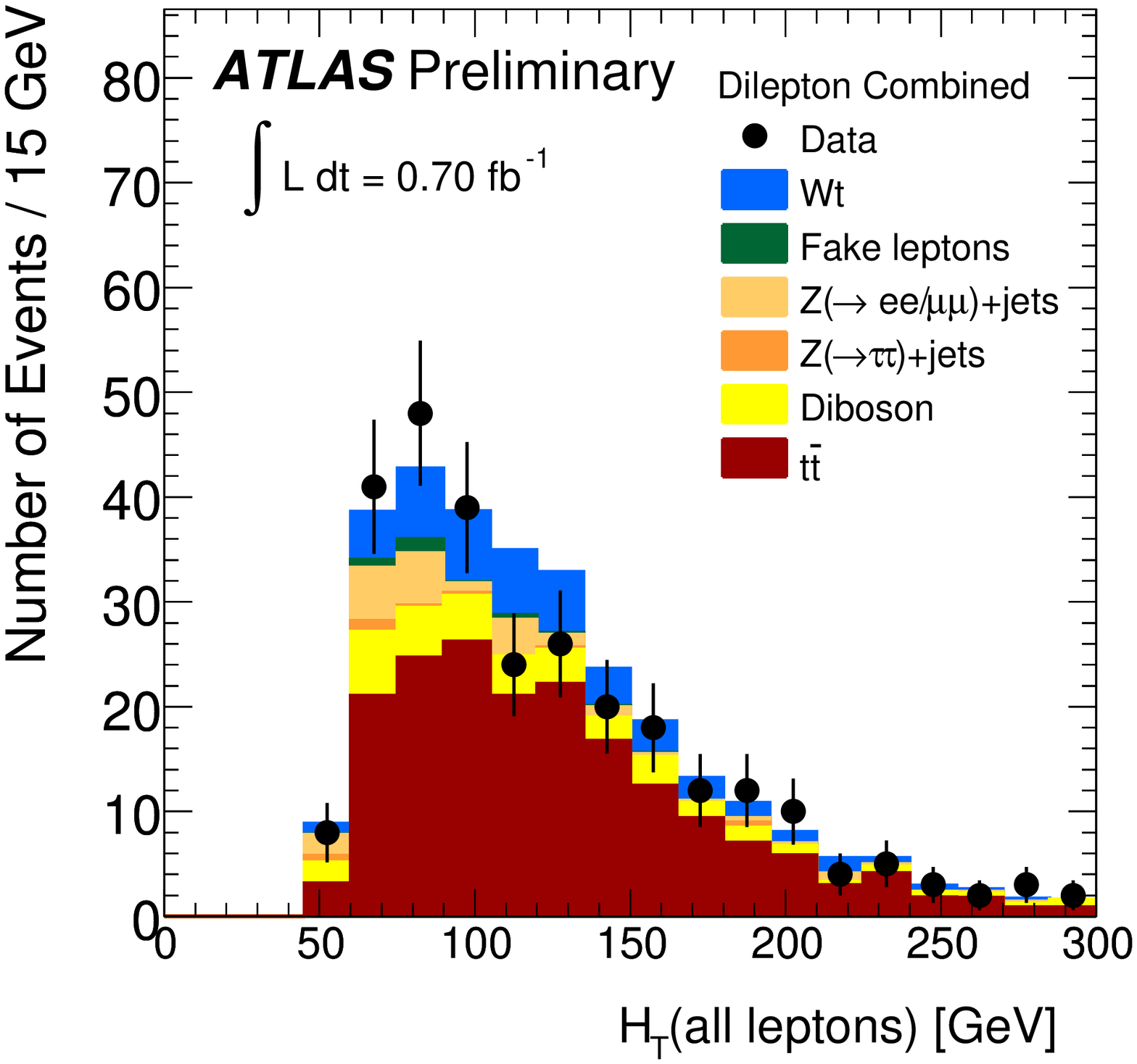}
\caption{Kinematic distributions after final event selection for the $\pt$ of the jet (left) and sum of the $\pt$ of the two leptons (right).}
\label{fig:WtCB}
\end{figure}

For the cut-based analysis there is only one cut, which requires exactly one jet.  Figure~\ref{fig:WtSel} shows the background removal accomplished by this selection, in particular the removal of \ttbar~ which tends to have a larger number of jets.  Distributions of the jet $\pt$ and the sum of the $\pt$ of the two leptons ($H_{\mathrm{T}}$(all leptons)) after this cut-based selection can be seen in Figure~\ref{fig:WtCB}.  The agreement between data and signal plus background model in these distributions is good.  The signal and background yields after the cut-based selection as well as the S/B ratio can been seen in Table~\ref{tab:SBWt}.  The e$\mu$ channel has the most events left after the selection (154 total events expected), and the S/B ratios are approximately 0.2 in all three channels.
 
\begin{table*}[t]
\begin{center}
\caption{Yields for signal and total signal plus background model, as well as the S/B ratio and data totals.  The cut-based analysis yields are determined after the application of the cut-based selection.}
\begin{tabular}{|l|c|c|c|}
\hline 
        &\textbf{ee} &\textbf{e$\mu$} &\textbf{$\mu\mu$} \\
\hline  single-top Wt  & 8.6 $\pm$ 1.6 &   26.6  $\pm$ 2.5  &   11.9  $\pm$ 1.7  \\
\hline  TOTAL Expected & 57  $\pm$ 7 &   154  $\pm$ 21 &   82  $\pm$ 10  \\
\hline  S/B            & 0.18 & 0.21 & 0.17  \\
 \hline DATA           & 62 & 152 & 73  \\
\hline 
 \end{tabular}
\label{tab:SBWt}
\end{center}
\end{table*}

The Wt cross-section and limit are extracted using a profile likelihood technique, the same method used in the $t$-channel cut-based analysis.  The resulting cross-section is $\sigma(pp\rightarrow Wt + X) = 14.4 ^{+5.3}_{-5.1} \mathrm{(stat)} ^{+9.7}_{-9.4} \mathrm{(syst)}$ pb.  The 95\% CL upper limit on the production of Wt events is $\sigma(pp\rightarrow Wt+X)~<~39.1$ pb observed, $\sigma(pp\rightarrow Wt+X)~<~40.6 $ pb expected.  The total expected uncertainty on the Wt limit is +77\% - 75\% with +68\% -66\% due to systematic uncertainties.  The data statistical uncertainty for this result is still large, +37\%, -35\%, so more data will improve this result.  However, there are also some large systematic uncertainties with values of about 30\%, particularly jet energy scale, jet energy resolution, and jet reconstruction efficiency.

\section{Conclusion}
ATLAS $t$-channel and Wt single-top cross-section results are estimated using 0.7$~\mathrm{fb}^{-1}$ of data taken with 7 TeV proton-proton collisions.  The $t$-channel cross-section is observed to be $\sigma_{t}= 90^{+32}_{-22}$ pb.  The Wt cross-section is found to be $\sigma(pp\rightarrow Wt + X) = 14.4 ^{+5.3}_{-5.1} \mathrm{(stat)} ^{+9.7}_{-9.4} \mathrm{(syst)}$ pb, corresponding to a 95\% CL limit of $\sigma(pp\rightarrow Wt+X)~<~39.1$ pb observed.

\bigskip % extra skip inserted

\end{document}